\documentclass[prd,nofootinbib,twocolumn,showpacs]{revtex4}
\usepackage{graphics} 
\usepackage{bm} 
 
\begin{document} 
 
\title{On formation of domain wall lattices} 
 
\author{Nuno D. Antunes} 
\affiliation{ 
Center for Theoretical Physics, University of Sussex, \\ 
Falmer, Brighton BN1 9WJ, United Kingdom.} 
 
\author{Levon Pogosian} 
\affiliation{ 
Theoretical Physics, The Blackett Laboratory, Imperial College,\\ 
Prince Consort Road, London SW7 2BZ, United Kingdom. } 
 
\author{Tanmay Vachaspati} 
\affiliation{
CERCA, Department of Physics, Case Western Reserve University, 
10900 Euclid Avenue, Cleveland, OH 44106-7079, USA.} 

\begin{abstract} 
     
We study the formation of domain walls in a phase transition in
which an $S_5\times Z_2$ symmetry is spontaneously broken to 
$S_3\times S_2$. In one compact spatial dimension we observe 
the formation of a stable domain wall lattice. In two spatial 
dimensions we find that the walls form a network with junctions, 
there being six walls to every junction. The network of domain 
walls evolves so that junctions annihilate anti-junctions. 
The final state of the evolution depends on the relative dimensions
of the simulation domain. In particular we never observe the formation
of a stable lattice of domain walls for the case of a square domain
but we do observe a lattice if one dimension is somewhat smaller than
the other. 
During the evolution, the total wall length in the network decays with
time as $t^{-0.71}$, 
as opposed to the usual $t^{-1}$ scaling typical of regular $Z_2$ 
networks.
\end{abstract}


\

\maketitle

\section{Introduction}
\label{intro}

Topological defects are routinely observed in condensed matter systems
and may be expected to form during phase transitions in the
early universe. Observational constraints on various types of defects
have already played a major role in the development of cosmology --
the idea of inflation was introduced in large part to solve the monopole
problem. Recently, much attention has been devoted to cosmological models 
motivated by String/M theory. Symmetry breaking patterns in these 
models can be quite complex with the production of topological defects 
being a generic phenomenon. 

It has been known for some time that, while topological considerations 
are sufficient for proving the existence of defects, the properties and 
interactions of defects depend on the details of the particular model. 
These issues were highlighted in earlier studies of domain walls in
an $SU(5)\times Z_2$ model. In contrast to the commonly studied
$\lambda \phi^4$ model, the vacuum manifold now consists of two disconnected
pieces, each of which is a 12 dimensional continuum. For topological 
reasons, kink solutions must exist. However, a 12 dimensional
continuum of possibilities exist when determining the boundary
conditions that the minimum energy kink solution satisfies. Topological
arguments alone are not sufficient in order to determine these 
boundary conditions and a more elaborate analysis is needed if 
one wants to find the kink solution \cite{PogVac01}. Only once 
the boundary conditions are known, can the explicit solution be
constructed and the properties worked out. In the case of domain
walls in $SU(N)\times Z_2$ models, this exercise has already yielded 
some surprises. For example, it was found that the symmetry group 
inside the core of a domain wall is generally smaller than that of 
the vacuum \cite{PogVac00,Vac01,PogVac01,Davetal02}. The interaction 
of kinks in these models shows that kinks and antikinks can repel as 
well as attract \cite{Pog02}. In Ref.~\cite{Vac03} it was argued that 
these features are in fact generic when large group symmetries 
are involved. Solutions of similar type were also discussed in 
\cite{Ton02}. 

The formation of kinks and domain walls in phase transitions has been 
studied in earlier work but most of this work only deals with the simplest 
of systems, such as the $Z_2$ kink in a $\lambda \phi^4$ model. 
Based on the expected initial density and scaling of these walls, 
cosmological implications have been derived. However, as we show here, 
the properties of a domain wall network in more complex models 
can be dramatically different from that of the wall network in the
$\lambda \phi^4$ model. Earlier work along these lines can be found
in Ref.~\cite{RydPreSpe90} where they consider the formation and
evolution of $Z_N$ walls, and in Refs.~\cite{KubIsiNam90,Kub92}
where the authors study the fate of walls in $O(N)$ motivated
models. 
 
An example of new physics that one can expect when considering domain 
wall formation in these more complex models was provided in \cite{PogVac03}. 
There it was shown that the $SU(N)\times Z_2$ models allow for the existence
of domain wall lattice solutions. The key ingredient is the repulsion 
between kinks and antikinks. The lattice is a periodic sequence of 
repelling walls and antiwalls that are parallel to each other.
It is more challenging to find a model in which the lattice is stable. 
In \cite{PogVac03}, a model in which $S_5\times Z_2$ ($S_n$ is the
permutation group of $n$ objects) breaks to $S_3\times S_2$ was used 
as an example of a model allowing for stable domain wall lattices. It 
was argued that in one spatial dimension, in the limit of many correlation 
domains, the probability of forming a lattice tends to unity. 

In this paper we follow on the work in \cite{PogVac03} and numerically 
investigate the formation of lattices during realistic phase transitions 
in (1+1) and (2+1) dimensions. Our results in 1+1 dimensions corroborate
the argument in \cite{PogVac03} and kink lattices are observed to form
with near certainty. The simulations in 2+1 dimensions, however, yield 
an unexpected feature that, with hindsight, might have been anticipated
from the discussion in Ref.~\cite{Davetal02}. Instead of forming 
a lattice, the domain walls form a network with junctions. Six walls 
meet at a junction and the relaxation of the network is controlled by 
the dynamics of the junctions. 
 
We find that this leads to considerable changes in the long time dynamical 
properties of the wall network. Whereas the total length of a $Z_2$ network 
of domain walls is expected to decay as $t^{-1}$, our junction dominated 
network has a lower scaling power: $t^{-0.71}$. Junction motion and 
junction/anti-junction annihilation processes clearly slowdown the long 
time evolution of the network. We did not observe the formation of a lattice 
of domain walls as the final evolution state in our simulations on a 
square spatial grid (with periodic boundary conditions). However, when the 
spatial grid is rectangular, with one dimension somewhat smaller (by roughly
a factor of 3) than the other, we did observe kink lattice formation, even
in 2+1 dimensions.

The paper is organized as follows. In Section ~\ref{model} we
introduce the model and review its features relevant for the
process of domain wall lattice formation.
Section ~\ref{formation} contains the simulation results in both
one and two spatial dimensions, including a discussion of the 
scaling regime of the wall network in the $2+1$ case.
Finally we discuss how these results may vary for more general
systems in different dimensions.
     
\section{Review of domain wall lattices} 
\label{model} 
 
In this section we review some of the results from 
Refs.~\cite{PogVac00,Vac01,PogVac01,Pog02,PogVac03} relevant to 
domain wall lattices. In \cite{PogVac03} $S_5 \times Z_2$ was used 
as an example of a model in which such lattice solutions are 
stable. However, the domain wall solutions themselves were 
identical to those in $SU(5)\times Z_2$ and we prefer, for clarity 
reasons, to use this model to describe their main properties 
\footnote{Most of the results in Section \ref{model} can 
be generalized to $SU(N)\times Z_2$ with $N>3$. The interested 
reader is referred to \cite{Vac01,PogVac01,Pog02,Vac03}.}. 
 
Consider an $SU(5)\times Z_2$ field theory described by a Lagrangian: 
\begin{equation} 
L = {\rm Tr} (\partial_\mu \Phi )^2 - V(\Phi ) 
\label{lagrangian} 
\end{equation} 
where $\Phi$ is an $SU(5)$ adjoint and $V(\Phi )$ is invariant 
under $SU(5) \times Z_2$. 
Let $V(\Phi )$ be such that the expectation value of $\Phi$ 
spontaneously breaks the symmetry down to 
$SU(3)\times SU(2)\times U(1)/Z_3\times Z_2$. 
We will choose $V(\Phi )$ to be a quartic polynomial: 
\begin{equation} 
V(\Phi ) = - m^2 {\rm Tr}[ \Phi ^2 ]+ h ( {\rm Tr}[\Phi ^2  ])^2 + 
      \lambda {\rm Tr}[\Phi ^4]  + V_0 
\label{quarticV} 
\end{equation} 
where $V_0$ is a constant chosen so that the minimum of the 
potential has $V=0$. The Lagrangian is symmetric under 
$\Phi \rightarrow -\Phi$ and it is the breaking of this 
$Z_2$ symmetry that gives rise to topological domain wall 
solutions. The desired symmetry breaking is achieved in the parameter range 
\begin{equation} 
{h \over \lambda} > - {7\over {30}} \ . 
\label{symmbreakparam} 
\end{equation} 
The vacuum expectation value (VEV), $\Phi_0$ is (up to any 
gauge rotation) 
\begin{equation} 
\Phi_0 = {\eta \over \sqrt{60}} {\rm diag}(2,2,2,-3,-3) 
\label{su5phi-} 
\end{equation} 
with $\eta \equiv {{m} / {\sqrt{\lambda '}}}$ and 
\begin{equation} 
\lambda ' \equiv h + {7\over {30}} \lambda \ . 
\label{lambdaprime} 
\end{equation} 
 
In Refs. \cite{PogVac00,Vac01,PogVac01} it was found that there are 
several kink solutions in this model corresponding to different 
choices of asymptotic field configurations. 
The necessary condition for the existence of a kink solution $\Phi_k(x)$, 
proved in Ref.~\cite{PogVac01}, is 
$[\Phi_k(x=\pm \infty),\Phi_k(x)]=0$. 
That is, the solution must commute with its asymptotic values. It was
also proved in Ref.~\cite{PogVac01} that,
when searching for $SU(5)$ kink solutions, one can work in the Cartan subalgebra 
of $SU(5)$, which is equivalent to restricting $\Phi_k(x)$ to a diagonal 
matrix form: 
\begin{equation} 
\Phi(x)=f_1(x)\lambda_3+f_2(x)\lambda_8+f_3(x)\tau_3+f_4(x)Y \ , 
\label{components} 
\end{equation} 
where $\lambda_3$, $\lambda_8$, $\tau_3$ and $Y$ are the 
diagonal generators of $SU(5)$: 
\begin{eqnarray} 
\lambda_3&=&\frac{1}{2} {\rm diag}(1,-1,0,0,0) \ , \nonumber \\ 
\lambda_8&=&\frac{1}{2\sqrt{3}} {\rm diag}(1,1,-2,0,0), 
\nonumber \\ 
\tau_3&=&\frac{1}{2} {\rm diag}(0,0,0,1,-1) \ , \nonumber \\ 
Y&=&\frac{1}{2\sqrt{15}} {\rm diag}(2,2,2,-3,-3) \ . 
\label{ymatrix} 
\end{eqnarray} 
 
As shown in Refs.~\cite{PogVac00,Vac01,PogVac01}, the kink solution with 
least energy is achieved if (up to global gauge rotations) 
$\Phi (-\infty) \equiv \Phi_- = \Phi_0$ and 
\begin{equation} 
\Phi (+\infty) \equiv \Phi_+ = 
- {\eta \over \sqrt{60}}{\rm diag}(2,-3,-3,2,2) 
\label{su5phi+} 
\end{equation} 
The minus sign in front of $\Phi_+$ in Eq.~(\ref{su5phi+}) puts $\Phi_+$ 
and $\Phi_-$ in disconnected parts of the vacuum 
manifold. Also, two blocks of entries 
of $\Phi_+$ are permuted with respect to those of 
$\Phi_-$. In other words, $\Phi_-$ and $-\Phi_+$ 
are related by a non-trivial gauge rotation. 
The kink solution (or, domain wall solution, in more 
than one dimension) can be written down 
explicitly in the case when $h/\lambda =-3/20$ 
\cite{PogVac00,Vac01}: 
\begin{equation} 
\Phi_k = {{1-\tanh(\sigma x)}\over 2} \Phi_- + 
                {{1+\tanh(\sigma x)}\over 2} \Phi_+ 
\label{qeq2solution} 
\end{equation} 
where $\sigma = m/\sqrt{2}$. 
{}For other values of the coupling constants, the 
solution has been found numerically \cite{PogVac00}. 
 
The topological charge of a kink can be defined as 
\begin{equation} 
Q = {{\sqrt{60}}\over \eta}(\Phi_R - \Phi_L) 
\end{equation} 
where $\Phi_R$ and $\Phi_L$ are the asymptotic 
values of the Higgs field to the right ($R$) and 
left ($L$) of the kink. (The rescaling has been 
done for convenience.) Then the charge of the kink 
in Eq. (\ref{qeq2solution}) is: 
\begin{equation} 
Q^{(1)} = {\rm diag}(-4,1,1,1,1) 
\label{Q1} 
\end{equation} 
Similarly, one can construct kinks with charge 
matrices $Q^{(i)}$ ($i=1,...,5$) which have $-4$ as 
the $ii$ entry and $+1$ in the remaining diagonal 
entries. Hence there are kink solutions with 5 
different topological charge matrices. Individually, 
the kinks can be gauge rotated into one another. But 
when two kinks are present, the different charges 
are physically relevant. This is most easily seen 
by noting that the interaction between a kink with 
charge $Q^{(i)}$ and an antikink with charge 
${\bar Q}^{(j)}= - Q^{(j)}$ is proportional to 
${\rm Tr}(Q^{(i)}{\bar Q}^{(j)})$ \cite{Pog02}. 
Then we have 
\begin{eqnarray} 
{\rm Tr}(Q^{(i)}{\bar Q}^{(j)}) 
               &=& -20 \ {\rm if} \ i=j \nonumber \\ 
               &=& +5  \ {\rm if} \ i \ne j 
\label{Qtraces} 
\end{eqnarray} 
The sign of the trace tells us if the force between the kink 
and antikink is attractive (minus) or repulsive (plus). 
Hence the force between a kink and an antikink with different 
orientations ($i\ne j$) is repulsive. This observation is key 
to the construction of kink lattices. 
 
A kink lattice is a periodic sequence of kinks with charges such 
that the nearest neighbour interactions are repulsive. One can write 
down a sequence of charges that can form a kink lattice \cite{PogVac03}: 
\begin{equation} 
... 
Q^{(1)}{\bar Q}^{(5)}Q^{(3)}{\bar Q}^{(1)}Q^{(5)}{\bar Q}^{(3)} 
... 
\label{minlatt} 
\end{equation} 
and the sequence just repeats itself. 
This sequence is the minimum sequence for which 
the nearest neighbour interactions are repulsive. 
Another way to write the kink sequence is to write it 
as a sequence of Higgs field expectation values. We write 
this sequence for the above minimal lattice: 
\begin{eqnarray} 
... &\rightarrow& +(2,2,2,-3,-3) 
              \rightarrow -(2,-3,-3,2,2) \nonumber \\ 
              &\rightarrow& +(-3,2,2,-3,2) 
              \rightarrow -(2,-3,2,2,-3) \nonumber \\ 
              &\rightarrow& +(2,2,-3,-3,2) 
              \rightarrow -(-3,-3,2,2,2) \nonumber \\ 
              &\rightarrow& +(2,2,2,-3,-3) 
                                 \rightarrow ... 
\label{Higgs sequence} 
\end{eqnarray} 
The minimal lattice of 6 kinks is not the only possibility. 
A sequence of 10 kinks in the $N=5$ case 
is aesthetic in the sense that it uses all the 5 different 
charge matrices democratically: 
\begin{equation} 
... 
Q^{(1)}{\bar Q}^{(5)}Q^{(3)}{\bar Q}^{(4)}Q^{(2)} 
{\bar Q}^{(1)}Q^{(5)}{\bar Q}^{(3)}Q^{(4)}{\bar Q}^{(2)} 
... 
\label{10latt} 
\end{equation} 
There can be longer sequences as well. 
 
A detailed stability analysis in Ref.~\cite{PogVac03} revealed that 
all domain wall lattices in $SU(5)\times Z_2$  are unstable. For example, 
the lattice in Eq. (\ref{minlatt}) has three unstable modes, corresponding to 
rotations in the 1-3, 1-5, 3-5 blocks. The instability comes from the fact 
that an isolated kink has zero modes corresponding to rotations 
in field space -- for example, a kink with charge $Q_1$ 
can be rotated into the kink with charge $Q_3$ without 
any cost in energy. When a kink of charge $Q_1$ is placed 
near an antikink of charge ${\bar Q}_3$, the zero mode 
becomes an unstable mode, making it favourable for $Q_1$ 
to rotate into $Q_3$ after which the kink and antikink 
can annihilate. 
 
To illustrate that stable lattices can exist, one can simply start 
with the model in which the zero modes are completely absent right 
from the start. As in Ref.~\cite{PogVac03}, let us consider the 
model of four real scalar fields $f_i$ ($i=1,..,4$), with 
\begin{equation} 
L = {1\over 2}\sum_{i=1}^4 (\partial_\mu f_i) ^2 
     + V(f_1,f_2,f_3,f_4) 
\label{s5lagrangian} 
\end{equation} 
and 
\begin{eqnarray} 
V &=& 
-{m^2\over 2}\sum_{i=1}^4 f_i^2 
+ {h \over 4}(\sum_{i=1}^4 f_i^2)^2 
+ {\lambda \over 8} \sum_{a=1}^3 f_a^4 
\nonumber \\ 
&+& {\lambda \over 4} \left[ 
{7\over 30} f_4^4 + f_1^2 f_2^2 \right] 
+ {\lambda \over {20}} [4(f_1^2 + f_2^2) + 9f_3^2] f_4^2 
\nonumber \\ 
&+& {\lambda \over \sqrt{5}}f_2 f_4 \left( f_1^2 
- {f_2^2 \over 3} \right) + {m^2 \over 4} \eta^2 
\label{potential} 
\end{eqnarray} 
The fields $f_i$ are defined as in Eq.~(\ref{components}) and 
this model has been obtained by substituting Eq.~(\ref{components}) 
into Eq.~(\ref{lagrangian}). 
This four field model does not have the continuous 
$SU(5)$ symmetry of the model in Eq. (\ref{lagrangian}). 
The only remnant of the $SU(5)$ symmetry corresponds 
to the permutation of the five diagonal entries of $\Phi$. 
In addition, the model also has the $Z_2$ symmetry under 
which $f_i \rightarrow -f_i$. Hence the model has an 
$S_5\times Z_2$ symmetry. 
 
A vacuum of the model is given by $f_1=0=f_2=f_3$ and 
$f_4 \ne 0$. This breaks the symmetry to
$S_3\times S_2$, corresponding to permutations of $\Phi$ 
in the $SU(3)$ and $SU(2)$ blocks. The vacuum manifold 
consists of $5!\times 2/3!\times 2! = 20$ discrete points. 
If we fix the vacua at $x=-\infty$, this implies that 
there are 20 kink solutions in the model. All these 
20 kink solutions have been described in Ref. \cite{PogVac01}. 
 
The construction of kink lattices proceeds exactly as in the 
$SU(5)$ case above because the off-diagonal components of $\Phi$ 
vanish there. Hence the $S_5\times Z_2$ model contains kink lattice 
solutions as well. Furthermore, these lattices are stable because the 
dangerous rotational perturbations are absent by the very construction 
of the model.  
 
\section{Lattice formation} 
\label{formation} 
 
There are more repelling kink-antikink pairs in $S_5$ than attracting 
ones. Hence, it is reasonable to expect that after the phase transition, 
all attracting walls will eventually annihilate and the remaining walls 
will be repelling.
 
The probability of forming a domain wall lattice in one spatial 
dimension for the $S_5$ model was estimated in \cite{PogVac03} to
be unity if the total number of kinks formed is large. Numerical 
simulations, presented below confirm this expectation. In two spatial
dimensions with periodic boundary conditions, our numerical 
results do not show lattice formation.
Instead the walls form a network with junctions that gradually
dilutes due to the annihilation of junctions.
 
\subsection{Numerical implementation}
\label{methods} 

In order to generate feasible initial conditions that  may lead to 
the formation of domain wall lattices, we use a Langevin type equation 
based on the Lagrangian Eq.~(\ref{s5lagrangian}). Each field will be 
propagated according to its usual equation of motion with additional 
dissipative and stochastic terms added:  
\begin{equation} 
\left( \partial_t^2 -\nabla^2 \right) f_i + \partial_i V 
 + D\, \partial_t f_i = \Gamma_i,  
\label{eom} 
\end{equation} 
$D$ is the dissipation constant and the stochastic 
force $\Gamma_i(x,t)$ is a Gaussian distributed field characterised by 
a temperature $T$: 
\begin{eqnarray} 
&& \langle \Gamma_i(x,t) \rangle = 0, \nonumber \\  
&& \langle \Gamma_i(x,t) \Gamma_j(x',t')\rangle= \frac{2 D}{T}  \delta_{ij}  
\delta (x-x')\delta (t-t')  
\label{noise} 
\end{eqnarray} 
The amplitude of the noise in Eq.~(\ref{noise}) is chosen so as to  
guarantee that independently of the initial field configuration and 
of the particular value of the dissipation, the system will always 
equilibrate towards a thermal distribution with temperature $T$. 
      
In both the $1D$ and $2D$ cases, we will start by evolving Eq.~(\ref{eom}) 
until thermal equilibrium is reached. At that point we {\it quench} the 
system to zero temperature by setting the stochastic term, $\Gamma_i$, 
in Eq.~(\ref{eom}) to zero. The fields will then 
settle towards the minima of the potential and a network of domain 
walls will form, separating regions where the fields were initialy  
uncorrelated. Note that the correlation length $\xi$ at thermal 
equilibrium depends on $T$ (at high temperatures, $\xi$ 
typically decreases with $T$). This  
allows us to have a degree of control over the correlation length 
of the fields before the quench, and hence over the number of  
independent domains that will form in the simulation box. This is
essential if we want to be certain to have a number of domains large 
enough to generate a stable lattice. 
 
The equations of motion were discretized using a standard leapfrog 
method and periodic boundary conditions were used. The model 
parameters  were set to $m=1/(2 \sqrt{6})$, 
$\lambda=1/2$ and $h=-3/40$. The lattice spacing was $\delta x=1.$ 
and $\delta t=.5$. Note that for the parameters above the wall core 
is resolved by more that 10 lattice points which should be accurate enough for 
the desired purposes. The value of the dissipation coefficient $D$  
does not influence the results during the stochastic stage of the 
simulation and we set it to $D=1.0$ to ensure rapid thermalization.  
In the next two sections we will discuss in more detail the role of
the dissipation after the quench. 
      
\subsection{Simulation in ($1+1$)D}
\label{1dim} 
 
\begin{figure}
\scalebox{0.50}{\includegraphics{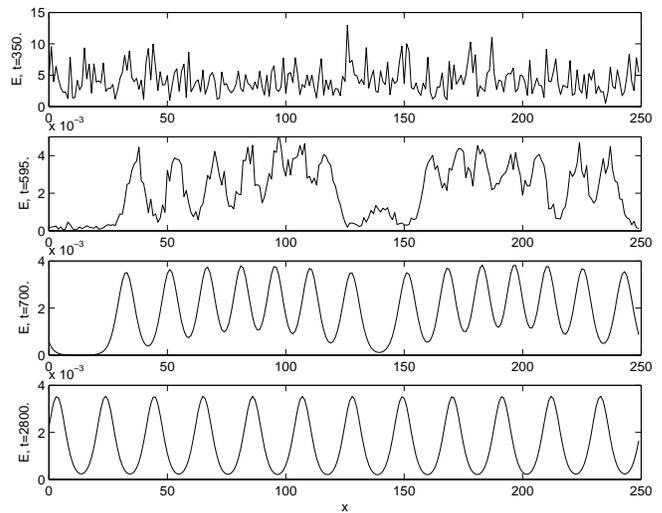}} 
\caption{Energy contour for a $1D$ system at different stages
of the evolution. The top plot shows the initial thermal configuration 
followed by a snapshot characteristic of early times after 
the quench to $T=0$. At later times the kink network evolves 
to form a stable wall lattice as shown in the bottom plot.} 
\label{fig1} 
\end{figure} 
     
Our goal here is to check whether a stable kink lattice forms in one 
spatial dimension as an initial ``hot'' configuration is quenched
to $T=0$. Applying the procedure detailed in the previous section to
the system, we confirm that this is in fact the case. In Fig.~\ref{fig1} 
we show  a series of snapshots of the energy profile of the fields 
for different times in a typical simulation run. The first configuration  
corresponds to the end of the thermalization phase, with large amplitude 
fluctuations and very uncorrelated fields. At this point we turn the 
stochastic term off and let the system evolve for a considerably long time. 
In this case we keep the dissipation high since we are not interested in 
the details of the dynamics of the system but rather in observing a stable 
lattice as the final outcome. As the energy is then dissipated, a 
recognisable pattern of ``proto-kinks'' starts to form, as can be seen 
in the second snapshot. The next energy profile shows that the
field has relaxed to a fully formed network of kinks. During further
evolution, neighbouring attractive kink-antikink pairs annihilate, leading 
to a stable lattice of equidistant domain walls. This final state is 
shown in the fourth plot, a succession of mutually repelling kinks and 
antikinks. The energy of these defects corresponds to the value predicted
analytically, and direct inspection of the values of the four fields $f_i$
confirms that, as expected, we are in presence of a genuine domain
wall lattice. 
 
%\begin{figure}
%\scalebox{0.50}{\includegraphics{evolution.ps}} 
%\caption{Energy contour for a $1D$ system at different stages
%of the evolution. The top plot shows the initial thermal configuration 
%followed by a snapshot characteristic of early times after 
%the quench to $T=0$. At later times the kink network evolves 
%to form a stable wall lattice as shown in the bottom plot.} 
%\label{fig1} 
%\end{figure} 
 
\subsection{Simulation in ($2+1$)D} 
\label{2dim} 
 
After having established that in one spatial dimension kink lattices
can form as a consequence of a phase transition, we will now look at
the evolution of the same model in the (2+1)D case. Here we expect
the late time dynamics of the system to be dominated by networks
of one-dimensional domain walls. 
  As a consequence of the symmetry content of the model, 
these networks will be considerably more complex than the ones based 
on ordinary $Z_2$ walls. In particular, we can expect walls to intersect 
and kink-antikink repulsion to play a role in the evolution.
     
Our first task is to find a method to identify the walls starting 
from the field 
values in the simulation lattice. To this end it is convenient  
to convert the fields $f_i(x)$ back into the original $SU(5)\times Z_2$ 
matrix form, $\Phi(x)$. For any type of kink in the model, there is at 
least one element $\Phi_{jj}$ of the diagonal of the field matrix that 
changes sign as one crosses the defect core \cite{PogVac00,Vac01,PogVac01}.
In the particular case of the lowest energy kinks, those described by the
charge matrices $Q^{(i)}$, it is easy to see that only
one of the $\Phi_{jj}$'s changes sign. For these cases, 
${\rm Tr}[\Phi^3]$ also vanishes at the defect's core. Since this does
not happen for any of the other kink types, we have a way of distinguishing
between the least energetic kinks and general unstable ones. We thus
measure at each time step the total number of zero crossings between all 
adjacent lattice points, of both the $\Phi_{jj}$'s and of ${\rm Tr}[\Phi^3]$. 
For most of the evolution the network consists predominantly of stable kinks
and the two quantities coincide, defining the total wall length in the
system.
     
The simulations follow the same pattern as before, starting
with a thermalization stage after which the stochastic term in
the equation of motion is set to zero. In this case, however we
are interested in studying the long time scaling behaviour of the
network. Since we are mostly concerned with the relativistic limit of the
theory the dissipation term must be set to zero during the scaling
period. Nevertheless, since the initial thermal configuration is very
energetic, we keep the dissipation high for some time after the
quench, before setting $D=0$  for the scaling regime. In 
this way, some of the excess energy is dissipated away, allowing the
fields to relax into a well formed network configuration that then
can start scaling.
     
 In all simulations, the model parameters and space and time
discretization steps used were the same as in the one dimensional 
runs. The fields were evolved in a $N=2500^2$ point grid.     
     
\begin{figure}
\scalebox{0.40}{\includegraphics{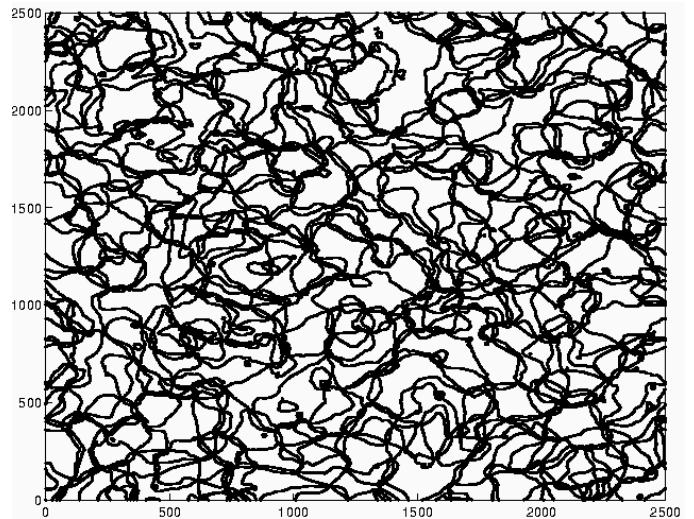}}
%\scalebox{0.50}{\includegraphics{walls_30.eps}}
\caption{High density wall network for an early
evolution time, $t=400$.}
\label{network1}
\end{figure}

\begin{figure}
\scalebox{0.40}{\includegraphics{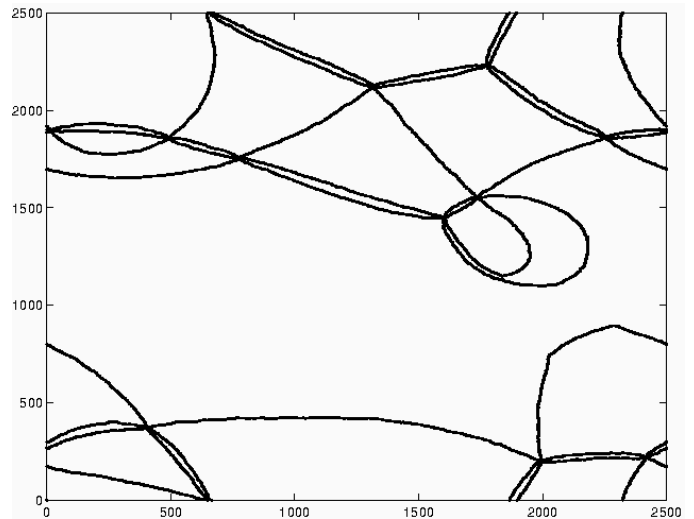}}
\caption{Late time network configuration $t=3200$.}
\label{network2}     
\end{figure}
   
In Fig.~\ref{network1} we can see a snapshot of the wall network
for an early time in the scaling period, $t=400$. The network is very dense
with a very high number of intersections. The walls were identified 
using the method discussed above. At this time all the kinks in the network
are stable and the number of zero
crossings for the $\Phi_{jj}$'s and for ${\rm Tr}[\Phi^3]$ coincide.
By direct observation of the field configurations, we confirmed
that higher energy kinks do exist for earlier times which quickly decay 
into stable ones. A secondary consequence of such decay processes
is that stable kinks forming in pairs from a single unstable kink 
tend to remain spatially correlated near junctions. Since the repulsive
force between kinks is exponentially
small, this pairing is relatively stable and can be observed for 
considerably late times (see  Fig.~\ref{network2}).

 For later times most
walls annihilate, loops form and decay into radiation and the
overall density decreases. Walls tend to get straighter as this is
energetically more favourable, though the time-scale relevant for
this process is clearly larger than the decay time of the network.
In Fig.~\ref{network2} we show the
network for a later time, $t=3200$. The reduction in density is
remarkable, with only a small number of intersecting walls
remaining. The way these intersect to form nodes is
quite interesting and we can see in particular  that around each
node there are always six incoming walls. This is related to the
fact that the minimal periodic wall lattice is composed
of six walls, as discussed in Section~\ref{model}. It is clear 
that for such a configuration to be stable for very long times,
the walls around the node must repel each other. Hence the
sequence of their charges must correspond to a stable wall
lattice pattern. For earlier times we observe nodes with other
numbers of incoming walls, though, as expected, never less
than six. High index nodes are rarer because they are
both less likely to form and more likely to cancel with other
nodes to form a minimal six wall configuration. Eventually all
nodes annihilate with anti-nodes with reversed ``vacuum orientation'',
and for very long times all walls disappear. In all the simulations
performed in a square lattice, the final state always had zero wall 
content and a two dimensional wall lattice never formed.
 Nevertheless, we observed that by taking one of the linear 
dimensions of the simulation domain $L_y$ to be smaller than the
other $L_x$, a regular
lattice of straight parallel walls can be obtained as the outcome 
of the evolution. 
Though more careful simulations need to be performed, our preliminary
results suggest that as the smaller grid dimension increases,
the final density of walls in the lattice becomes smaller. When
$L_y/L_x$, the 
ratio of the two dimensions becomes larger than roughly $1/3$, no 
lattice is formed.
This result can be better understood by imagining      
 the limit when one of the torus
 dimensions (say $L_y$) goes to zero and the 1D case is recovered.
 In this situation, the final state of the evolution should be
 a series of parallel walls crossing the torus in the $y$-direction.
 As $L_y$ increases, we can expected that some of these walls
 will be able to ``explore'' the torus in both the $x$ and
 $y$-directions. Some of
 these will not cross the torus, getting linked to other walls
 and forming a complex network. As the network
 evolves, wall nodes will annihilate, some of these annihilations
 further decreasing the number of walls that cross the box in
 the $y$-direction. As a result, the final state of the system
 will still be a set of straight repelling parallel walls in the
 $y$-direction, but with a lower density. It is easy to imagine that
 as $L_y$ increases and approaches $L_x$, the final density will decrease,
 up to the point where no lattice will form at all.    
     The fact that the final state of the transition
depends on the geometry of the system is unexpected and may have 
implications in cosmological settings in models with extra dimensions. 
Still, a better understanding of this phenomenon and the
mechanisms behind it is needed before further conclusions can be drawn.
  
\begin{figure}
\scalebox{0.50}{\includegraphics{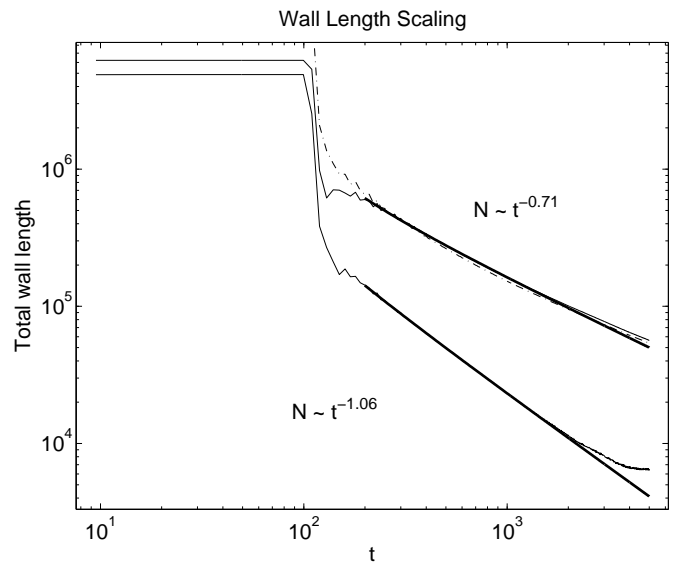}}
\caption{Wall length versus time for both the $S_5\times Z_2$ theory
(top curve) and the single field $Z_2$ case (bottom curve). The dashed
curve corresponds to the total $S_5\times Z_2$ wall length measured 
by counting zeros of the diagonal elements of $\Phi(x)$. Except
for very early times this coincides with the number of zeros of
${\rm Tr}[\Phi^3]$ (solid curve). The results
shown are averages of $25$ independent realizations. In all cases,
for $t<100$
the system is evolved with both non-zero dissipation and thermal
noise. At $t=100$ we quench the system to zero temperature but the
dissipation term is kept until $t=125$ to eliminate 
excess energy. For $t>125$ we set $D=0$. The power-law fits (bold
lines) to both curves  were taken between $t=200$ and $t=1350$. For
later times (larger than half the simulation box length) finite size
effects become significant.}
\label{scalling}
\end{figure}

Another indication that the dynamics of this system differs
fundamentaly from regular defect networks can be obtained by studying
 its scaling properties. 
In Fig.~\ref{scalling} we show a $log-log$ plot of the total wall
length versus time. For comparison we also include a plot of
the scaling of a regular $Z_2$ domain wall network. This was 
obtained by evolving a scalar $\lambda \phi^4$ theory in parallel
with the $S_5\times Z_2$ system. In both cases after the initial
thermalization and dissipation periods, the domain wall network 
enters a scaling regime where the time evolution of the
total wall length is well described by a power-law $N\sim t^{-\alpha}$.
For the $Z_2$ theory we find $\alpha=1.06(0.06)$ 
 which is in reasonable
agreement with both previous theoretical and numerical predictions
of $\alpha=1.$ (see \cite{GarHin02} and references therein). This is 
to be compared with the result for the $S_5\times Z_2$ walls.
In this case we find $\alpha=0.71(0.02)$, a considerably lower result.
Clearly, the dynamics of the $S_5\times Z_2$ network is fundamentally 
different
from the $Z_2$ case. This is not surprising, taking into account that
the process of node/anti-node annihilation must play
an important role in the dynamics of the system. That the overall
effect is a slowdown in the scaling, could also be expected on the
basis that the six-wall junctions feel a force that is predominantly
isotropic. This suggests that systems allowing for stable nodes
with higher number of crossing walls and hence more isotropic, would
be likely to display even lower scaling powers. In the limit of
infinite number of incoming walls per node the system would eventually
become static.

\section{Conclusions}
\label{conclusions}
 
We have numerically studied phase transitions in a model 
with $S_5\times Z_2$ symmetry breaking down to $S_3\times S_2$
in one and two spatial dimensions. In one dimension, as expected, 
we find that a stable domain wall lattice forms. In two dimension,
we observed the formation of a complicated network of domain walls 
and domain wall junctions. At every junction six domain walls
are present. With time, the junctions move and annihilate, and
the network coarsens. This feature of the network dynamics leads
to a slowdown of the scaling regime. 
We found that the total length of domain wall
in the network scales as $t^{-0.71}$, in contrast to
the $t^{-1.0}$  fall off
expected for $Z_2$ domain walls in a $\lambda \phi^4$ model.

We have not studied the phase transition in three dimensions 
since the problem then becomes computationally very intensive.
Even then our results can be extrapolated to three dimensions,
allowing us to anticipate certain behavior. In three dimensions
we expect the wall junctions to be one dimensional -- somewhat
like strings. The network of walls will coarsen as the strings
come together and annihilate. This is very reminiscent of a system 
of cosmic strings in which each string then gets connected to six 
domain walls and the evolution of the networks should be similar
\cite{KubIsiNam90,Kub92,RydPreSpe90,VilShe94}. The new feature
in the present case is the repulsion between walls and antiwalls
and this could lead to lattice formation at very late times.

Perhaps the most important application of our study is to the
$SU(5)\times Z_2$ case as this is a realistic particle physics
model. As we have noted in Sec.~\ref{model}, the domain walls
in the $SU(5)$ model are identical to those in the $S_5$ model.
The difference is only in their stability properties. Hence
we might expect some of the features of the domain wall networks
in the $S_5$ case to carry over to the $SU(5)$ case. Exactly
how much similarity will survive is hard to predict and may
depend on model parameters. This remains an important problem
to explore.

\begin{acknowledgments}
We would like to thank Mark Hindmarsh and Arthur Lue for useful comments
and suggestions and Raymond Volkas for pointing out an error in the text
of the paper. 
We thank the organizers of the ESF COSLAB School in Cracow where a 
part of this project was completed. TV was supported by DOE grant 
number DEFG0295ER40898 at CWRU. NDA was supported by a PPARC
post-doctoral fellowship.
The numerical simulations  were carried on the COSMOS Origin2000 supercomputer
supported by Silicon Graphics, HEFCE and PPARC.
 
\end{acknowledgments}

\end{document}